# Understanding High Temperature Superconductors: Progress and Prospects

David Pines

Center for Nonlinear Studies, LANL, and Physics Dept. and Science and Technology Center for Superconductivity, University of Illinois at Urbana-Champaign, 1110 West Green Street, Urbana, IL 61801-3080.

I review progress in measurements of the dynamic spin susceptibility in the normal state which yield a new phase diagram and discuss microscopic calculations which yield qualitative, and in many cases, quantitative agreement with the measured changes in the quasiparticle, transport, magnetotransport, and optical properties of the cuprate superconductors as one varies doping and temperature provided one describes the systems as nearly anti-ferromagnetic Fermi liquids in which the effective magnetic interaction between planar quasiparticles mirrors the dynamic spin susceptibility measured in NMR and INS experiments. Together with the demonstration that the NAFL pairing potential leads inexorably to a $d_{x^2-y^2}$ pairing state, this work provides a "proof of concept" for the NAFL description of high $T_c$ materials. I review Eliashberg calculations of the mean-field behavior found in overdoped systems and discuss the extent to which the crossovers to pseudoscaling and pseudogap behavior found in the effective magnetic interaction and quasiparticle behavior in the optimally doped and underdoped systems may be derived microscopically. I conclude with a tentative scenario for the dependence of $T_c$ on doping level and imperfections in different systems.

## 1. INTRODUCTION

Understanding high temperature superconductors means, first of all, answering the following questions: What is the nature and physical origin of their anomalous normal state behavior? How is the normal state best characterized? What is the mechanism for high $T_c$? What is the pairing state? Thanks to a large number of interrelated experimental and theoretical investigations carried out during the past three years, we have, as a community, made considerable progress toward obtaining the answers to these key questions and in what follows I present some highlights of that progress.

My answers may be simply stated. There is an intimate relationship between the anomalous evolution of planar quasiparticle spectra with temperature and doping and the anomalous spin and charge behavior measured in the normal state of optimally doped and underdoped systems: all these phenomena originate in a highly anisotropic magnetic interaction between the almost localized planar quasiparticles. The close approach to anti-ferromagnetism leads to three distinct kinds of normal state magnetic behavior: mean field, pseudo-scaling, and pseudogap; counterparts are found in the quasiparticle spectra, transport, magnetotransport, and optical behavior. The normal state may thus be characterized as a nearly antiferromagnetic Fermi liquid, with the dominant contribution to quasiparticle interaction coming from spin-fluctuation exchange, and so mirroring the momentum and frequency dependence of the dynamical spin susceptibility, $\chi(\mathbf{q},\omega)$, measured in NMR and INS experiments. Because $\chi(\mathbf{q},\omega)$ peaks sharply at wave vectors $\mathbf{Q}_i$, in the vicinity of $\mathbf{Q} \equiv (\pi,\pi)$, there are two distinct groups of quasiparticles: *hot quasiparticles*, located near the singular points on the Fermi surface which can be connected by $\mathbf{Q}_i$, interact strongly and display highly anomalous (non-Landau Fermi liquid) behavior; *cold quasiparticles* are sufficiently far away from singluar points that their behavior is closer to that of a strongly coupled Landau Fermi liquid. Recent microscopic calculations based on the NAFL description of the doping and temperature dependence of the transport, magnetotransport, and optical properties of both overdoped and underdoped systems yield good qualitative, and in many cases, quantitative agreement with experiment [1], while pseudogap behavior, and more generally the magnetic crossovers found in even optimally doped systems, is associated with the hot quasiparticles [2]. The highly anisotropic quasiparticle interaction found in NAFL's leads inexorably to a transition to a superconducting state with $d_{x^2-y^2}$ pairing [3], and I shall present arguments which suggest that the doping dependence of $T_c$ originates in a competition between the spin-fluctuation-induced pairing potential and the quasiparticle pseudogap. The enhancement of $T_c$ on going from 1-2-3 systems to Tl- and Hg-based systems is explained, in the NAFL scenario, by differences in their spin fluctuation spectra [4], while the markedly different influence of the planar impurities, Zn and Ni, on the quasiparticle pseudogap in underdoped systems, and on $T_c$ in optimally doped systems, can be quantitatively understood using the NAFL description [3][5], and provides strong support for the magnetic origin of high $T_c$, and of pseudogap behavior in the normal state.

## 2. SPIN-FLUCTUATION EXCITATIONS AND TEMPERATURE CROSSOVERS: A NEW PHASE DIAGRAM.

A quantitative fit to existing NMR and INS experiments on the 1-2-3 and 2-1-4 systems may be obtained with a phenomenological expression, $\chi_{NAFL}(\mathbf{q},\omega)$ for the dynamical spin susceptibility which takes into account the strong temperature dependent antiferromagnetic correlations between the itinerant $Cu^{2+}$ spins [6]. $\chi_{NAFL}$ possesses in general four incommensurate peaks at wave vectors, $\mathbf{Q_i}$, in the vicinity of $(\pi,\pi)$ and for $(\mathbf{Q_i}-\mathbf{q}) \lesssim 1/\xi$ takes the form [7]

$$\chi_{NAFL}(\mathbf{q},\omega) = \frac{1}{4}\sum_i \frac{\chi_{\mathbf{Q_i}}}{1+(\mathbf{Q_i}-\mathbf{q})^2\xi^2 - i\omega/\omega_{SF}} \quad (1)$$

elsewhere. $\chi(\mathbf{q},\omega)$ takes a more Fermi liquid-like form,

$$\chi_{FL}(\mathbf{q},\omega) = \frac{\chi_q}{1-i\omega/\Gamma_q} \cong \frac{\chi_0(T)}{1-i\omega/\Gamma_0} \quad (2)$$

where $\chi_Q \equiv \alpha\xi^2$ is the static spin susceptibility, $\xi$ the AF correlation length, $\omega_{SF} \ll (\Gamma_q,\Gamma_0)$ describes NAFL relaxational behavior, $\chi_0(T)$ is the uniform spin susceptibility, and $\Gamma_q \sim \Gamma_0$ is the effective bandwidth or Fermi energy. The $^{63}Cu$ spin-lattice relaxation time and spin echo decay experiments discussed in Refs. (2) and [7] show that for magnetically overdoped systems one finds mean-field behavior, $\omega_{SF} \sim \xi^{-2} \sim a + bT$, in the normal state. For optimally doped and underdoped systems, such behavior is found only above a characteristic temperature, $T_{cr}$, close to the temperature at which $\chi_0(T)$ exhibits a maximum in the normal state of these systems. Barzykin and I [7] argued that the crossover in system behavior at $T_{cr}$ occurs when the AF correlations reach a critical strength, $\xi_{cr} \sim 2a$; thus $\xi(T_{cr}) \cong 2a$ determines $T_{cr}$, a criterion which was confirmed by the subsequent INS experiments on 2-1-4 discussed in Ref. (8) and the $^{63}T_{2G}$ measurements on $YBa_2Cu_4O_8$ reported in Ref. (9). As T decreases below $T_{cr}$, $(d\chi_0/dT) > 0$, and $\omega_{SF}$ exhibits pseudoscaling behavior, $\omega_{SF} = \hat{c}/\xi = a' + b'T$, until a second crossover temperature, $T_*$ is reached. $T_*$ marks the onset of pseudogap behavior, characterized by $\xi \sim$ constant, and a rapid increase in $^{63}T_1T$ and $\omega_{SF}$ for $T_c \leq T \leq T_*$ [7]. The new phase diagram obtained from an analysis of NMR and INS experiments on the 1-2-3 and 2-1-4 systems is shown in Fig. 1. Note that $T_{cr}$ increases rapidly as the doping level falls below that which is optimal for $T_c$ while $T_*$ increases more slowly. For optimally doped 1-2-3, one finds $T_{cr} \sim 150K$, $T_* \sim 120K$, while for $YBa_2Cu_4O_8$, Curro *et al.* [9] have shown that $T_{cr} \cong 500K$, $T_* \sim 200K$. For $La_{1.85}Sr_{0.15}CuO_4$, one has $T_{cr} \sim 300K$ [8], while Yasuoka [10] finds $T_* \sim 200K$. Note, too, the considerable variation from system to system in the strength of AF correlations at $T_c$; for $YBa_2Cu_3O_7$ one finds $\xi(T_c) = 2.2a$, while for $La_{1.86}Sr_{0.14}CuO_4$, one has $\xi(T_c) = 7.6a$ [8]. Equally striking is the increase in $\xi(T_c)$ within a given system as one goes to sub-optimal doping levels; as might be anticipated from Fig. 1, one finds $\xi(T_c) = 5.5a$ for $YBa_2Cu_4O_8$.

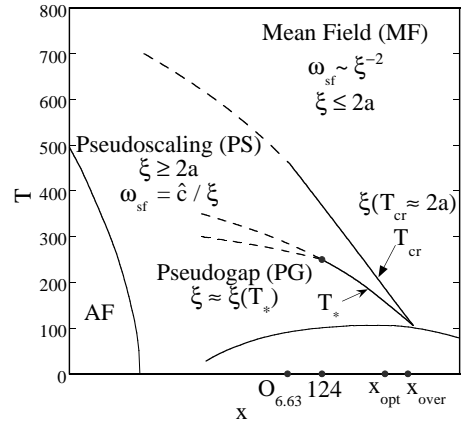

Fig. 1 Phase diagram, showing three distinct normal state phases for magnetically underdoped members of the 1-2-3 system. Extrapolations of the crossover temperatures, $T_{cr}$ and $T_*$, to lower doping levels are denoted by dotted lines. Comparable phase diagrams have been established for the 2-1-4, Bi-based, Hg-based and Tl-based systems.

Note, too, that from a magnetic perspective, optimally doped systems are, in fact, underdoped; only when one goes to substantially higher doping concentrations (e.g., $La_{1.76}Sr_{0.24}CuO_4$) does one reach the magnetically overdoped regime, where $\chi_{NAFL}$ displays mean-field behavior, $(d\chi_0/dT) < 0$ down to $T_c$ and $\xi(T_c) \leq 2$.

For magnetically overdoped systems, and in the mean-field regime for magnetically underdoped systems, microscopic strong coupling Eliashberg calculations of the kind carried out for $YBa_2Cu_3O_7$ [3], may be expected to provide a quantitative account of normal state properties. The non-linear feedback of changes in the quasiparticle spectrum on the effective quasiparticle interaction in the NAFL description,

$$V_{eff}(\mathbf{q},\omega) = g^2 V_{NAFL}(\mathbf{q},\omega) \qquad (3)$$

which determines their behavior brings about both the comparatively slow increase, $\xi(T) \sim (a+bT)^{-1/2}$, found above $T_{cr}$, and a moderate increase in $\chi_0(T)$ as T decreases, while the recent work of Chubukov and Monthoux [11] shows that vertex corrections are comparatively modest, are maximal for the hot quasiparticles, and act to enhance the pairing potential for superconductivity. However, as discussed in Ref. (2), and as the criterion, $\xi(T_{cr}) = 2$ implies, for magnetically underdoped systems below $T_{cr}$ the spectra of the hot quasiparticles [which must be located near $(0,\pm\pi)$ and $(\pm\pi,0)$] begins to be modified by the strong AF correlations in such a way that one finds pseudo-scaling behavior for $\omega_{SF}$ between $T_{cr}$ and $T_*$; the pseudogap behavior below $T_*$, represents the "end-point" of the impact of the strong AF correlations on quasiparticle properties. This gapping of the hot quasiparticles at $T_*$ is seen directly in the rapid increase in $\omega_{SF}$ below $T_*$, in the BSCCO ARPES experiments [12], and in the specific heat and uniform susceptibility measurements [13] discussed at this conference.

The magnetic origin of the quasiparticle pseudogap has a number of observable consequences. First, only hot quasiparticles should become gapped. Second, both the gap magnitude and its onset temperature $T_*$, should be unrelated to $T_c$. Third, once in place, at temperatures slightly below $T_*$, the hot quasiparticle gap may be expected to be independent of T, *and should not be influenced by the superconducting transition*. Fourth, since the gapped hot quasiparticles do not participate in the superconducting transition, for magnetically underdoped systems, the superfluid density, $\rho_s$, will be distinctly less than the number of quasiparticles (~1-x), one estimates from the area of the large Fermi surface expected above $T_{cr}$ (or $T_*$), and is plausibly ~x, the hole fraction.

## 3. HOT AND COLD QUASIPARTICLES: TRANSPORT, MAGNETOTRANSPORT, AND OPTICAL PROPERTIES

Microscopic calculations of the properties of hot and cold quasiparticles in a NAFL reveal dramatically different behavior, and show how these differences give rise to the measured anomalous dependence of transport, magneto-transport, and optical properties as one varies the hole concentration and temperature [1]. One finds quite generally that hot quasiparticle scattering rates are considerably larger than those for cold quasiparticles with

$$\frac{1}{\tau_{hot}} \sim T\xi \quad (4)$$

while the cold quasiparticle lifetime depends on its distance, $\Delta k$, from the nearest hot spot;

$$\frac{1}{\tau_{cold}} \sim \frac{T^2}{(\Delta k)^3 \xi^2 \omega_{SF}} \qquad T \ll \frac{\omega_{SF}(\Delta k)^2 \xi^2}{\pi} \quad (5a)$$

$$\frac{1}{\tau_{cold}} \sim \frac{T}{(\Delta k)} \qquad T \gg \frac{\omega_{SF}(\Delta k)^2 \xi^2}{\pi} \quad (5b)$$

The main contribution to the longitudinal conductivity, $\sigma_{xx}$, comes from the cold regions of the FS; up to logarithmic corrections, one finds [1]

$$\rho_{xx} \sim \frac{T^2}{T_0 + T} \frac{1}{(\Delta k)_{max}} \quad (6)$$

where the crossover temperature, $T_0$, is

$$T_0 = (\Delta k_{max}) \xi^2 \frac{\omega_{SF}}{2\pi} \quad (7)$$

and $\Delta k_{max} \sim 1$ for a large FS. For magnetically underdoped systems, Fermi surface evolution leads to $\Delta k_{max} \lesssim 0.3$; hence for a wide choice of parameters $T_0 < T_*$, and one finds a linear in T resistivity down to $T_*$, while the rapid increase in $T_0$ below $T_*$ leads to the crossover to $T^2$ behavior seen in $YBa_2Cu_3O_{6.63}$ below 160K and in $La_{1.9}Sr_{0.1}CuO_4$ below 150K. The cold quasiparticles likewise dominate the Hall conductivity, $\sigma_{xy}$, so that one finds, for $T \lesssim 2\pi T_0$,

$$ctn\theta_H \sim \frac{T^2}{T_0} \quad (8)$$

and one recovers the famous $T^2$ dependence of $ctn \theta_H$ for optimally doped and overdoped systems. In Ref. (1) it is shown how the highly anisotropic hot and cold quasiparticle mean-free paths enable one to extract $\tau_{cold}$ and $\tau_{hot}$ from measurements of $\sigma_{xx}$ and $\sigma_{xy}$; in so doing, one finds for $YBa_2Cu_3O_7$, $YBa_2Cu_4O_8$, $YBa_2Cu_3O_{6.63}$, as well as for $La_{1.9}Sr_{0.1}CuO_4$ and $La_{1.85}Sr_{0.15}CuO_4$, quantitative agreement with the theoretical expressions, Eqs. (4) and (5).

On extending these results to finite frequencies, $\omega$, the terms involving both $\omega$ and $\tau$ cancel; Eqns. (5) then become

$$\frac{1}{\tau_k(\omega, T)} \sim \frac{1}{\Delta k} \left[ \frac{\pi T^2}{T + T_0} + \frac{\omega^2}{\omega + \pi T_0} \right]. \quad (9)$$

At typical frequencies and temperatures one then recovers the marginal Fermi liquid result,

$$\frac{1}{\tau_k(\omega, T)} = A\omega + B\pi T. \quad (10)$$

One finds thereby agreement with optical experiments on materials as disparate as single layer Tl 2201 ($T_c$ = 23K) and $YBa_2Cu_3O_7$ [1].

## 4. WHAT DETERMINES $T_c$?

As discussed in detail in Refs. (3), in a NAFL the peaking of the highly anisotropic quasiparticle interaction at wave vectors near $(\pi,\pi)$ leads inexorably to a $d_{x^2-y^2}$ pairing state, while strong coupling (quasiparticle lifetime) corrections do not prevent one from obtaining a $T_c \sim 90K$ with para-meters appropriate to YBa$_2$Cu$_3$O$_7$ and a coupling strength which yields a $\rho_{xx}$(T) comparable to that found experimentally. This "proof of concept" for the NAFL description led Monthoux and me to predict a $d_{x^2-y^2}$ state for all cuprate superconductors, at a time when only NMR experiments supported this pairing assignment. As we heard at Grenoble, and in still greater detail here, our prediction has now been supported by essentially all experiments which are sensitive to the symmetry of the order parameter.

For the optimally doped and overdoped systems for which our Eliashberg calculations are applicable, we found that

$$T_c \sim \omega_{SF}\, \xi^2 \tag{11}$$

so that for comparable values of $\xi(T_c)$, $T_c$ should scale with $\omega_{SF} \sim {}^{63}T_1T$. This scaling relation has been invoked to explain both the substantially higher transition temperatures found in the Tl- and Hg-based systems, where one finds values of $\xi(T_c)$ comparable to those of YBa$_2$Cu$_3$O$_7$, but values of $\omega_{SF}(T_c)$ some 50% larger [4], and the variation of $T_c$ with Ni doping measured in YBa$_2$Cu$_3$O$_7$ [5]. On examining the sensitivity of our results to $\xi(T_c)$, we found that for $\xi \lesssim 1$, $T_c$ plummetted to zero, so that, as might be expected of a mechanism which depends on a near approach to AF behavior, overdoping, which weakens the AF correlations, brings out the reduction in $T_c$ seen experimentally.

On the other hand, how can we account for the observed maximum in $T_c(x)$, since in the pseudoscaling regime, Eq. (11) leads one to expect monotonic behavior, with $T_c \sim \xi$? The answer of course lies in the pseudogap, which by removing hot quasiparticles from the superconducting arena, weakens substantially the overall NAFL pairing potential. On this scenario, as one enters the magnetically underdoped regime, initially the increasing strength of the AF correlations wins out over the pseudogap, but below the optimal doping concentration, the pseudogap wins out, and $T_c$ falls off as x decreases. Obviously, further calculations are required to support this scenario, in which the dimensionless coupling constant, N(0)g, is assumed to vary weakly with hole concentration.

Support for the above scenario comes from experiments on the influence of the planar impurities, Ni, and Zn, on $T_c$. For YBa$_2$Cu$_3$O$_7$ Monthoux and I argued that since Zn does not possess a spin which can align with neighboring Cu$^{2+}$ spins, it will change the magnetic pairing potential and so act as a "superunitary" scatterer, while Ni, which enters the plane in such a way as to look like a spin 1/2 to its neighboring Cu$^{2+}$ spins, will be a subunitary scatterer. In this way we could explain the fact that Zn is some three times more effective than Ni in suppressing $T_c$ for YBa$_2$Cu$_3$O$_7$, and argued that this effect serves as a "smoking gun" for the magnetic mechanism [3]. But how then to explain the fact that both Ni and Zn exert a somewhat larger influence on $T_c$ in YBa$_2$Cu$_4$O$_8$? The answer lies in the fact that the pseudogap is well developed in this system so that the density of states at $T_c$ is significantly reduced; hence the increased effectiveness of both Ni and Zn impurities in reducing $T_c$ in underdoped systems. Because it changes the pairing potential as well, Zn continues to be a more effective pair-breaker than Ni.

## 5. SOME NEXT STEPS

Although we possess a good understanding of optimally doped and overdoped systems, much work remains to be done before we possess a comparable understanding of the normal state properties and superconducting transition temperature of the magnetically underdoped systems. Some next steps include the development of a microscopic description of hot spot, pseudoscaling, and pseudogap behavior, which makes evident the interplay between the influences of Fermi surface evolution and pseudogap formation on the measured quantities, $\omega_{SF}$ and $\chi_Q$, and establishes a quantitative relationship between $\rho_s$, $\chi_0$, and the doping level. Experiments on the same samples of BSCCO and other materials are needed to verify the correctness of the two-gap scenario proposed in Sec. 2 and to establish the connection between features measured in ARPES and INS experiments and crossovers measured in NMR experiments. With regard to microscopic theories, at present we know what doesn't work: neither one band Hubbard models with U or nearest neighbor exchange or the t-J model yield a quantitative account of the pseudoscaling and pseudogap behavior measured in underdoped materials, while the predictions of the non-linear sigma model plus holes are inconsistent with the doping dependence of pseudoscaling behavior. What might work is far less clear. A one or three band Hubbard model with a non-local restoring force which takes into account local SDW formation at $\xi \geq 2a$ is a promising candidate, but much work will be required before its viability can be assessed.

It gives me pleasure to acknowledge stimulating conversations on these and related topics with my collaborators, A. Balatsky, V. Barzykin, A. Chubukov, P. Monthoux, J. Schmalian, and B. Stojkovic and with my UIUC experimental colleagues, G. Blumberg, D. Ginsberg, L. Greene, M. Klein, C. P. Slichter, R. Stern, and D. Van Harlingen, and to thank CNLS and CMS at Los Alamos, and STCS NSF grant DMR91-20000 at UIUC for the support provided to this work.

REFERENCES

1. B. Stojkovic and D. Pines, Phys. Rev. B, in the press (April 1, 1997 issue); in preparation.
2. A. Chubukov, D. Pines and B. Stojkovic, J. Phys. Cond. Matt. **8** (1996) 1; D. Pines, Z. Phys. B, in the press (1997).
3. P. Monthoux and D. Pines, Phys. Rev. B **47** (1993) 6069; Phys. Rev. B **49** (1994) 4261; D. Pines, Physica C **235-236** (1994) 113; D. Pines and P. Monthoux, J. Phys. Chem. Solids **56** (1995) 1651.
4. G-q. Zheng *et al.*, J. Phys. Soc. Jpn. **64** (1995) 2524; K. Magishi *et al.*, J. Phys. Sco. Jpn **64** (1995) 4561; M. Julien *et al.*, Phys. Rev. Lett. **76** (1996) 4238.
5. Y. Tokunaga *et al.*, these proceedings.
6. A. Millis, H. Monien and D. Pines, Phys. Rev. B **42** (1990) 167.
7. V. Barzykin and D. Pines, Phys. Rev. B **52** (1995) 13585.
8. Y. Zha, V. Barzykin and D. Pines, Phys. Rev. B **54** (1996) 7567.
9. C. P. Slichter *et al.*, Phil. Mag. **74** (1996) 545; N. Curro et al., Phys. Rev. B, in the press (1997).
10. H. Yasuoka, these proceedings.
11. A. V. Chubukov and D. Morr, Physics Reports, in the press (1997); P. Monthoux, Phys. Rev. B, in the press (1997).
12. Z-X. Shen, these proceedings; H. Ding, *ibid.*
13. J. Loram, these proceedings; J. Tallon, *ibid.*